# The gate tunable 2D *pn* junction driven out-of-equilibrium

*Ferney A. Chaves and David Jiménez

*Departament d'Enginyeria Electrònica, Escola d'Enginyeria,
Universitat Autònoma de Barcelona, Campus UAB, 08193 Bellaterra, Spain.
*ferneyalveiro.chaves@uab.cat*

**Abstract**

We have investigated the electrostatics and electronic transport of the gate tunable 2D *pn* junction by implementing a comprehensive physics-based simulator that self-consistently solves the 2D Poisson's equation coupled to the drift-diffusion current and continuity equations. The simulator considers the strong influence of the out-of-plane electric field through the gate dielectric and the presence of interface states. The impact of parameters such as gate capacitance, energy gap and interface trap states density have been considered to model properties such as the depletion width, rectification factor and depletion and diffusion capacitances. The present work opens the door to a wider exploration of potential advantages that gate tunable 2D *pn* junctions could bring in terms of figures of merit.

## I. INTRODUCTION

Semiconductor *pn* junctions (PNJs) are basic building blocks for many applications both in electronics and optoelectronics. Doping elements intended to be incorporated as chemical impurities into the semiconductor have been key to the development of this kind of applications. However, with the appearance of devices with size of a few nanometers, the application of a conventional impurity doping scheme faces challenges such as difficulty to form junctions with extremely high doping gradients and control of random doping fluctuations to avoid device-to-device variability [1-4]. The accurate control of doping type, density, and spatial distribution in nanostructures is also challenging because of their complex growth and geometrical constrains [5-8]. For any nanoscale device, the requirement of a high carrier density along with unintentional ionized dopants in the active region points toward impurity free doping solutions.

Even though chemically doped 2D layered materials based PNJs have been already demonstrated [9-16], the chemical doping is not straightforward for materials such as graphene, phosphorene, silicene, black phosphorous or transition metal dichalcogenides (TMDs), [17,18] making difficult the fabrication of controlled high doped junctions with low defect density so limiting their performance. Consequently, various approaches have been recently proposed to control the electron and hole concentrations in 2D materials based PNJs (2D PNJs) by means other than chemical doping (CD). Some of these approaches take advantage of the electrostatic interaction between the semiconductor and a different material at the interface to modify the carrier density [4,19-21]. Other proposals make use of external electric fields produced by a single gate or two separated gates to control the semiconductor carrier density. Those approaches are termed as "electrostatic doping" (ED). ED potentially offers ultrasharp junctions with well-controlled carrier density profile and a reduced defect density. These features make ED an attractive alternative to CD junctions [11,12,22-25]. From the theoretical point of view, the focus has been put on the understanding of both the electrostatics and transport properties of chemically doped 2D PNJ [26,27,36,28-35]. However, despite the experimental efforts to explore the electronic and optoelectronic performances of gate tunable (GT) 2D PNJs [37-41], theoretical research of these devices is still lacking behind.

In this paper we have developed a physics-based numerical simulator to study the electrostatics and electric transport of GT 2D lateral PNJs (GT 2D LPNJs). Our simulator self-consistently solves the 2D Poisson's equation coupled to the drift-diffusion current and continuity equation, providing a tool for the numerical computation of the electrostatics and transport properties, in an effort to provide basic understanding of the experimental measurements. The simulation domain has been defined according to Fig. 1, which is intended for the study of the device intrinsic behavior. Extrinsic effects such as parasitic capacitances and contact resistances are beyond of the scope of this work and therefore are not considered. Our present work can be also useful for the assessment of future compact models oriented to circuit simulation and as a tool for the design of new device concepts based on 2D PNJs.

In Section II we have presented a model for the electrostatics of the GT 2D LPNJ. Sections III and IV discuss both the electronic transport and capacitances of the device, respectively. In Section V the proposed model has been benchmarked with experimental data and finally, in Section VI, we have drawn the main conclusions of our work.

## II. DEVICE ELECTROSTATICS

In this section we discuss the electrostatics of the GT 2D LPNJ schematically shown in Fig. 1, which sets the basis for the later formulation of the electron transport model. The ultrathin semiconducting material under the electric control of two back gate metal electrodes, separated by a gap of length $L_{gap}$, makes possible to induce both hole and electron carriers depending on the selected gate voltages $V_{g1}$ and $V_{g2}$, so different types of junctions can be created pp, nn, pn or np. For instance, to prepare a symmetric GT 2D LPNJ, the gate bias should be selected as $-V_{g1} = V_{g2} = V_g$. To bias the device out-of-equilibrium one might select $V_1 = -V_2 = V/2$, where $V(= V_1 - V_2)$ is the applied bias between the cathode and anode electrodes. The band diagram represented at the top of Fig. 1 corresponds to the symmetric case at thermal equilibrium where hole (p-type) and electron (n-type) carriers are induced by negative and positive gate voltages, respectively. The 2D semiconductor sheet is assumed to make ohmic contacts with the metal electrodes and the carrier transport is produced between both edges upon application of a bias. The direction of the current transport defines the longitudinal direction ($x$), the transversal direction ($z$) goes from the back gates to the semiconductor sheet and the $y$-direction goes across the width of the device ($W$), where $W$ is large enough so the device can be considered uniform along this direction.

At thermal equilibrium ($V = 0$), part of the electrostatically induced majority carriers move by diffusion to the adjacent region until an equilibrium state is reached. Thus, the condition of zero net electron and hole current requires that the Fermi level is constant throughout the semiconductor. Owing to this fact, an in-plane electric field arises, the energy bands are bent and a region depleted of carriers, characterized by a width $W_d$, is formed near the junction ($x = 0$) separating both $p$- and $n$-type induced charge regions. When an intrinsic 2D semiconductor is considered, the depletion region does not have any fixed charge and the induced charge regions have a non-zero total charge, unlike CD 3D and 2D PNJs in their quasi-neutral regions.

### 1. 1D model of the metal-oxide-semiconductor (MOS) structure

In this Subsection analytical expressions for the gate tunable Fermi level at the semiconductor layer are provided. As represented in Fig 2a, the cross section of a GT 2D LPNJ consists of a MOS like structure. As for the analysis we have considered the Fermi level at the induced charge region far away from the depletion region. Let us consider a point inside the induced charge region with an electrochemical potential $V'$ respect to an arbitrary reference. At this point, the vertical electric field produced by the gate biased at $V_g$, induces a carrier charge density $Q_{sc}$ in the semiconductor with energy gap $E_g$, related to the electrostatic potential $\phi_o$. A relation between $V_g$ and $\phi_o$ can be gotten by considering the band diagram of the MOS structure, shown in Fig. 2b, resulting from a vertical cut along the GT 2D LPNJ in Fig. 2a. From the charge conservation law:

$$Q_m + Q_{sc} + Q_{dop} + Q_{it} = 0 \quad (1)$$

where $Q_m$ is the gate charge density, $Q_{sc} = q(p - n)$, $Q_{dop}$ is the charge density corresponding to possible chemical doping and $Q_{it}$ is the interface trapped charge density between the semiconductor and the gate dielectric. The electron ($n$) and hole ($p$) carrier densities are given by $n_s \log(1 + \exp([q(\phi_o - V') - E_0]/kT))$ and $n_s \log(1 + \exp([-q(\phi_o - V') - E_0]/kT))$, respectively, with $n_s = g_{2D}kT$, where $g_{2D}$ is the band-edge density of states (DOS), $E_0 = E_g/2$, $k$ the Boltzmann constant and $T$ the absolute temperature. On the other hand, the voltage Kirchhoff's law along the structure holds:

$$W_m - qV_{ox} = \chi + E_0 - q\phi_o + qV_g \quad (2)$$

where $W_m$ is the gate metal workfunction, $\chi$ is the semiconductor electron affinity and $V_{ox}$ is the potential drop across the dielectric layer. For example, in the n-type region, the combination of Eqs. 1 and 2 results in:

$$[-\phi_o + V'_g]C_g - qn + Q_{dop} + Q_{it} = 0 \quad (3)$$

where $V'_g = V_g - (W_m - \chi - E_0)/q$ is defined as the gate voltage overdrive and $C_g = \epsilon/T_{ox}$ is the gate capacitance per unit area, being $\epsilon$ the insulator dielectric constant and $T_{ox}$ the insulator thickness. In our model $Q_{it}$ is calculated following the procedure developed in [42]

$$Q_{it} = -q \int_{E_V}^{E_F} N_{it,A} dE + q \int_{E_F}^{E_C} N_{it,D} dE \quad (4)$$

where $E_F$, $E_C$ and $E_V$ refer to the semiconductor Fermi energy level, the bottom of the conduction band, and the top of the valence band, respectively; and $N_{it,A}$ ($N_{it,D}$) is the acceptor (donor) interface states density. After some algebra, and considering $V' = 0$ for the equilibrium case, Eq. 4 results in:

$$Q_{it} = -q^2 N_{it}\phi_o + q(N_{it,D} - N_{it,A})E_0 \quad (5)$$

where $N_{it} = N_{it,A} + N_{it,D}$. By replacing Eq. 5 into Eq. 3 a non-linear equation for $\phi_o$ must be solved. The induced doping level $E_F - E_i = q\phi_o$ as a function of $V_g'$ for wide range of values of $E_g$ (0.6 - 1.8 eV) is shown in Fig. 3a, where the vertical axis is normalized to $E_0$. An ideal device free of interface traps ($Q_{it} = 0$) and chemical doping ($Q_{dop} = 0$) has been considered. In this work, operation of the device at room temperature is assumed. $E_F - E_i$ quickly increases in a linear way for small values of $V_g'$ followed by a non-linear behavior at larger $V_g'$. Moreover, the threshold gate voltage $V_{th}$ at which that transition occurs increases with $E_g$. As shown in Fig. 3b, considering realistic values of $N_{it}$ at the semiconductor-dielectric interface, that increases the value of $V_{th}$. For simplicity, $N_{it,A} = N_{it,D}$ is assumed, so that the neutral level is located at the midgap [42]. Here, we have considered $E_g = 1.8$ eV (such as the monolayer MoS$_2$) and $T_{ox} = 300$ nm. As an exemplary case, for $N_{it} = 2.5 \times 10^4$ eV$^{-1}$μm$^{-2}$, the inset of Fig. 3b shows $N_{in} = p - n$ versus $V_g'$ in both linear and logarithmic scales, exhibiting $V_{th} \sim 26\ V$ and an exponential trend in the subthreshold region followed by a linear trend in the conduction region. Also, for the sake of comparison, the results obtained from the numerical solution of the 2D Poisson's equation (described later) at thermal equilibrium are represented by solid black dots in the main panel of Fig. 3, showing a perfect fit with the 1D model described by Eq. 3. Despite Eq. 3 is a transcendental equation without analytical solution, closed expressions for both $\phi_o$ and $N_{in}$ can be obtained in both the subthreshold and conduction regions. Firstly, for the subthreshold region:

$$\phi_o = (E_F - E_i)/q = V_g'/\gamma \quad (6a)$$

$$N_{in} \approx n_i \exp(qV_g'/\gamma kT) \quad (6b)$$

where $\gamma = 1 + q^2 N_{it}/C_g$ and $n_i \approx n_s \exp(-E_0/kT)$ is the semiconductor intrinsic carrier density. Then, for the conduction region the following approximated expressions hold:

$$\phi_o \approx \frac{E_0}{q} + \frac{kT}{q} \log\left(e^{\frac{(V_g' - V_{th})C_g}{qn_s}} - 1\right) \quad (7a)$$

$$N_{in} \approx (V_g' - V_{th})C_g/q \quad (7b)$$

$V_{th}$ can be estimated as 90% of $\gamma E_0/q$ to get a good fit between Eq. 7 and the numerical solution of Eq. 3. Importantly, Eq. 7 keeps valid for electrically degenerated semiconductors.

Finally, the impact of $C_g$ on the Fermi level, for a semiconductor with $N_{it} = 0$ and $E_g = 1.8$ eV, has been plotted in Fig. 3c. For a fixed $V_g'$, the larger $C_g$ the higher the value of the Fermi level and therefore the larger $N_{in}$. Also, the larger $C_g$ is the smaller $V_g'$ at which the device starts to be degenerated.

The results of this section are useful for a better understanding of the depletion width ($W_d$) behavior, especially in the presence of interface trap states, as well as the device out-of-equilibrium properties, described in next sections.

## 2. Depletion width model

The GT 2D LPNJ is particularly useful for optoelectronics because the existence of a controllable built-in electric field in the depletion layer that drives the photogenerated e-h pairs towards the device terminals, contributing to a sizeable photocurrent [37]. Even more, they have the advantage that $W_d$ can be modulated either by $V_g'$, by $L_{gap}$ or even by $T_{ox}$ as it will be shown later. In this section we have benchmarked $W_d$ of GT 2D LPNJs at thermal equilibrium against both CD 3D-PNJs and CD 2D LPNJs.

As reported elsewhere [42], the depletion width for a symmetric CD 3D PNJ at equilibrium is given by:

$$W_{d,3D} = \sqrt{\frac{4\epsilon_{sc}\phi_{bi}}{qN_{3D}}} \quad (8)$$

where $\epsilon_{sc}$ is the bulk semiconductor dielectric permittivity, $\phi_{bi}$ the built-in potential of the junction and $qN_{3D}$ the doping concentration. On the other hand, the effective depletion width for symmetric CD 2D LPNJs has been derived as: [26,27,34]

$$W_{d,2D} = \frac{\pi^2 \epsilon_{eff} \phi_{bi}}{4GqN_{2D}} \quad (9)$$

where $\epsilon_{eff}$ is the effective dielectric constant, G is the Catalan constant and $N_{2D}$ is the doping density.

In order to determine $W_d$ in GT 2D LPNJs and its dependence on parameters as $V_g'$, $T_{ox}$, $\epsilon$ and $N_{it}$, we firstly solve numerically the 2D non-linear Poisson's equation at thermal equilibrium conditions to find out the in-plane electrostatic potential profile $\phi(x) = \phi(x, T_{ox})$ and the charge density profile $Q_{sc}(x)$ along the $x$ direction. By considering an infinitely thin semiconductor the 2D Poisson's equation can be reduced to the Laplace's equation, which is solved inside the simulation domain shown in Fig. 1, where Dirichlet boundary conditions are defined at the gate contact boundaries, followed by the boundary condition $\partial \phi / \partial z = -\sigma(\phi)/\epsilon$ at the semiconductor plane, and Neumann homogeneous boundary conditions at the rest of the domain boundaries. Then, analogously to our previous works ([34,35]), $W_d$ is numerically determined according to the criterium $Q_{sc}(x)$ reaching 43.6% of the induced charge $qN_{in}$ at each side of induced charge regions of the junction. For the sake of simplicity, we have assumed a symmetric GT 2D LPNJ, although the proposed model can also deal with asymmetric devices. Fig. 4a shows the dependence of $W_d$ with $V_g'$ for devices with different values of $T_{ox}$, once fixed the parameters $\epsilon = 3.9\epsilon_o$, $N_{it} = 0$, $E_g = 1.8$ eV and $L_{gap} \to 0$. Generally, $W_d$ exhibits a monotonically decreasing trend with $V_g'$ due to the increasing of induced doping $N_{in}$ like in CD PNJs. The inset in Fig. 4a shows a perfect linear dependence of $W_d$ on $T_{ox}$ with slope $\sim 0.70$. There, for every $T_{ox}$, $W_d$ has been computed at $V_g'$ such that $\phi_{bi}/V_g' = 0.08$, where $\phi_{bi} = 2\phi_o$. The observed behavior suggests that $W_d$ exhibits a simple linear dependence with both $T_{ox}$ and $\phi_{bi}/V_g'$. By plotting $W_d$ versus $\phi_{bi}/V_g'$ for several $T_{ox}$, as shown in Fig. 4b, the simulations effectively predict a linear trend, which fits pretty well to the simple formula:

$$W_d = 0.6 T_{ox}\left[2\frac{\phi_{bi}}{V_g'} + 1\right], \quad (10)$$

revealing a non-explicit dependence on $\epsilon$, in remarkable contrast to the CD case according to Eqs. 8-9. The range of $V_g'$ considered in Fig. 4b belongs to the conduction region. The weak dependence of $W_d$ on $\epsilon$, as shown in Fig. 4c, comes through $C_g$ according to Eq. 7a.

Next, we have explored the effect of $L_{gap}$ on $W_d$ for a device with gate biases $V_{g1} = -V_{g2} = -10$ V. As expected, when $L_{gap} \ll W_d$ produces an almost constant $W_d$. Further increase of $L_{gap}$ results in a slope equal to one and $W_d$ is then completely determined by $L_{gap}$.

Finally, the effect of non-idealities has been considered. Specifically, the dependence of $W_d$ with $N_{it}$ is shown in Fig. 4e for the symmetric device with $T_{ox} = 300$nm operated at $V_g' = 20$V. Interestingly two different ranges can be distinguished. In the range from 0 until $\sim 2 \times 10^4$ eV$^{-1}$ μm$^{-2}$ the device works in the conduction region, and $W_d$ monotonically increases with $N_{it}$ like in the CD 2D LPNJ case [35]. As for the range $N_{it} > 2 \times 10^4$ eV$^{-1}$ μm$^{-2}$ the device works in the subthreshold region, and $W_d$ does decrease instead. Equivalently, a peaked behavior of $W_d$ vs $V_g'$ is also expected for a fixed value of $N_{it}(\neq 0)$. For instance, observing the curve corresponding to $N_{it} = 2.5 \times 10^4$eV$^{-1}$ μm$^{-2}$ of Fig. 3b, $W_d$ vs $V_g'$ is expected to increase from zero until $\sim 25$ V in the subthreshold region and decrease from $\sim 25$ V onwards in the conduction region the same way as for the case with no traps included. The difference in the behavior of $W_d$ with and without interface trap-states relies on the ratio between $qN_{in}$ and $Q_{it}$. In the conduction region ($V_g' > V_{th}$) the large density of induced carriers screens the effect of trapped states and therefore $W_d$ behaves like in the case with no traps included. In contrast, when the device is operated in the subthreshold region, the induced carriers are very few and the trapped charges dominate the electrostatics.

### III. DEVICE ELECTRONIC TRANSPORT

In this Section, we deal with the out-of-equilibrium behavior of the GT 2D LPNJ. Here we have considered the drift-diffusion (DD) as the dominant electronic transport mechanism. Other mechanism has been suggested in 2D FETs to explain the experimental data such as the hopping transport [43–45]. Those works suggest that charge transport in the subthreshold regime is consistent with a variable range hopping (VRH) model although in the linear regime the DD model does better. The electrostatics of the here studied device must be solved across the $xz$-plane and characterized by the 2D electrostatic potential distribution $\phi(x, z; V_1, V_2, V_{g1}, V_{g2})$. $V_1$ and $V_2$ are the voltages at anode/cathode terminals ($V_2$ taken as the voltage reference), thus the bias applied to the device can be written as $V = V_1 - V_2$. Consideration of the out-of-equilibrium regime requires account the coupling between the electrostatics and the charge transport equations. In our work, we have assumed a 1D DD mechanism acting as the driving force. In fact, the transport equations involve a system of four nonlinear first-order differential equations, two of them for the electron and hole quasi-Fermi potentials $V_{n(p)}(x)$ and two for the electron and hole current densities $J_{n(p)}$. The equations contain key physical parameters such as the electron (hole) mobility $\mu_{n(p)}$, the minority electron (hole) lifetime $\tau_{n(p)}$, the energy gap $E_g$, and the operation temperature $T$. The in-plane 1D potential $\phi(x) = \phi(x, T_{ox}; V_1, V_2, V_{g1}, V_{g2})$ and the quasi-Fermi potentials $V_{n(p)}$ are related to the local intrinsic energy $E_i$ and electron (hole) Fermi energy $E_{Fn(p)}$, through the relation $E_i(x) = -q\phi(x)$ and $E_{Fn(p)}(x) = -qV_{n(p)}(x)$, respectively. Generation and recombination processes, mediated by capture and emission of carriers at trap centers located in the band gap of the semiconductor, are accounted by means of the net recombination rate $U = G - R$ according to the Shockley–Read–Hall model [42]. A complete description of the mathematical procedure and the numerical method used to solve self-consistently the equations describing the electrostatics and electronic transport can be found in our previous work [35].

*J-V* characteristics for a prototype (free of interface trapped charge) device driven out-of-equilibrium are shown in Fig. 5a for several $V_g'$'s. We have assumed a device in both *pn* and *np* configurations after selecting $V_{g1} < 0$, $V_{g2} > 0$ and $V_{g1} > 0$, $V_{g2} < 0$, respectively. Considered transport related parameters are $\mu_n = \mu_p = 5$ cm²/Vs and $\tau_n = \tau_p = 100$ ps. As expected, the current displays a rectifying behavior for both type of junctions. Importantly, at forward bias both the *np-* and *pn-*junctions exhibit an ideality factor $\eta$ roughly equal to 2, suggesting that recombination, rather than diffusion current, dominates the forward current, consistently with observations reported in [37]. By following the same procedure reported elsewhere for CD 3D PNJs, it is possible to find a simple expression for the recombination current density $J_R$ at $V > kT/q$ [46]:

$$J_R = q \int_0^{W_d} U dx \approx \sqrt{\frac{\pi}{2}} \frac{kT n_i}{\tau F_o} \exp\left(\frac{qV}{2kT}\right) \quad (11)$$

where $F_o$ is the in-plane electric field at the location of maximum recombination $U$, which in turn depends on $V$. The factor 2 in the exponential term corresponds to the ideality factor. For the sake of simplicity, we have assumed in Eq. 11 that $\tau = \tau_n = \tau_p$ and $L_{gap} \to 0$. As shown in the upper inset of Fig. 5b, the profile of the electric field $F(x)$ inside the depletion region is non-linear, in contrast with the CD 3D PNJ (linear trend) and CD 2D LPNJ (1/x trend) [35]. The lower inset of Fig. 5b shows the dependence of $F_o$ on $V$. The main panel of Fig. 5b shows the *J-V* characteristics computed at forward bias coming from numerical simulations, plotted together with the recombination current from Eq. 11, which give support to the fact that the recombination current dominates the forward current. To get a perfect fit with Eq. 11, we have multiplied the numerically obtained current density by 1.3, which can be justified by the effect produced by a non-zero $L_{gap}$ (assumed to be 0.02 µm as the value 0 µm cannot be physically implemented).

Next, the effect of $L_{gap}$ on the transport characteristics has been considered. The results are shown in Fig. 5c. The larger $L_{gap}$ is the higher the current, coming from the increase of $W_d$ which reduces the value of $F_o$. As a complementary information, the inset of Fig. 5c. shows the 2D electrostatic potential for $L_{gap} = 0.02$ µm and 1.2 µm.

For realistic simulations of the device, the presence of interface states within the forbidden gap must be considered. The impact of interface states on the electrical properties of 2D FETs has been addressed in some studies elsewhere [47–50] but there are a lack of studies on their impact on GT 2D LPNJs. In this sense, we address how the diode rectification is affected by $N_{it}$.

Starting from the same simple model given by Eq. 4, but now considering the quasi-Fermi potentials $V_n$ and $V_p$ for electrons and holes, respectively, the following expression for $Q_{it}$ can be obtained:

$$Q_{it} = -q^2(N_{it,A}V_n + N_{it,D}V_p) - q^2(N_{it,A} + N_{it,D})\phi + q(N_{it,D} - N_{it,A})E_0 \quad (12)$$

which must be included in the 2D Poisson's equation to solve the out-of-equilibrium 2D electrostatics. The effect of $N_{it}$ on the *J-V* characteristics is shown in Fig. 6a, where $-V_{g1} = V_{g2} = 20$V and $N_{it,A} = N_{it,D}$ have been assumed. In general, the currents for both forward and reverse bias increase with $N_{it}$ because of the screening of the induced charge produced by the trapped charge. As shown in the inset of Fig. 6a, the presence of trapped charges lowers the density of induced charge carriers inside the induced charge regions (p and n) and thus the built-in potential $\phi_{bi}$ of the device is also reduced, resulting in a smaller potential barrier for electrons and holes that eases the carrier transport. Experimental evidence of the reduction of the carrier density can be found in [4]. As shown in Fig. 6b, the current has a monotonically decreasing behavior with $V_g'$ at a reverse bias of 0.5V for the values of $N_{it}$ here assumed, while at the forward bias 0.5 V, the current exhibits a maximum at $V_g' \sim 13$V for $N_{it} = 2.5 \times 10^4$ eV⁻¹µm⁻² and at $V_g' \sim 25$V for $N_{it} = 5 \times 10^4$ eV⁻¹µm⁻². These maxima reflect the double behavior of $W_d$ with $V_g'$ as discussed in the previous section.

Next, we define the Rectification Factor (RF) as $RF = J(V = 0.5V)/J(V = -0.5V)$. As observed in Fig. 6c, RF parameter is quite sensitive to $V_g'$. Interestingly, in this case the presence of interface states in the 2D semiconductor/insulator interface contribute to get better RF values. However, owing to the increasing of the channel resistance (decreasing of induced carrier density) when $N_{it}$ increases, or due to the contact resistance in real devices, the RF might be degraded.

## IV. DEVICE CAPACITANCES

In this Section we have analyzed the depletion and diffusion capacitances of the GT 2D LPNJ. For the sake of simplicity, we have focused on the symmetric PN case.

Since in a GT 2D LPNJ made of intrinsic semiconductor there are no fixed charges in the depleted region, unlike the CD 3D and 2D cases, the calculation of the depletion capacitance relies on the following expression:

$$C_{dp} = \frac{d}{dV} \int_0^{W_d/2} q(p-n) dx, \qquad (13)$$

which, in principle, must be numerically solved. However, inside of the n-type side of the depletion region $n(x) \gg p(x)$ and $n(x)$ can be approximated to an average electron density $\bar{N}$. Thus, in complete analogy with the CD PNJ, a simple expression for the depletion capacitance can be obtained from Eq. 13:

$$C_{dp} = -\frac{d}{dV}\left(q\bar{N}\frac{W_d}{2}\right) = \frac{q\bar{N}}{2}\frac{d}{dV}\left(0.6 T_{ox}\left[2\frac{\phi_{bi}-V}{V_g'}+1\right]\right) = 0.6 q\bar{N}\frac{T_{ox}}{V_g'} \qquad (14)$$

By considering $\bar{N}$ as a constant fraction of $N_{in}$ such that $\bar{N} = N_{in}/\alpha$, and using Eq. 7b applied to devices with no interfacial states, $C_{dp}$ can be expressed as:

$$C_{dp} = 0.6\epsilon/\alpha \qquad (15)$$

which is only dependent of $\epsilon$ as for CD 2D LPNJs [35], but not on $T_{ox}$ or $V_g'$. Fig. 7a shows a comparison between Eq. 15 and the values obtained from the simulations, where $\alpha = 5.3$ has been chosen to get the best fit. In Fig. 7b we have represented the numerical calculations of $C_{dp}$ for several values of $\epsilon$ as a function of $T_{ox}$ compared with Eq. 15, revealing its independence with $T_{ox}$. Fig. 7c shows the behavior of $C_{dp}$ with $V_g'$ (assuming $-V_{g1} = V_{g2} = V_g'$) from numerical results, which exhibits a weak dependence not considered in the simple model given by Eq. 15.

As for the diffusion capacitance, it has been numerically computed as

$$C_{df} = 2\frac{d}{dV}\int_{W_d/2}^{\infty} q\left(p - \frac{n_i^2}{N_{in}}\right) dx \qquad (16)$$

where we have considered the excess of holes (minority carrier) in the n-type induced charge region. However, it is possible to use the same procedure as the one used for CD 3D PNJs [42], which allows to derive a simple expression for the low frequency diffusion capacitance:

$$C_{df} = \frac{q^2 n_i^2 L}{kT N_{in}} e^{qV/kT} \qquad (17)$$

where $L$ is the electron/hole diffusion length. Finally, Fig. 7d shows both the numerically computed depletion and diffusion capacitances as a function of $V$ for a prototype GT 2D LPNJ biased at $V_g' = 20$ V. The device parameters $T_{ox} = 300$nm, and $E_g = 1.8$ eV and $E_g = 1.1$ eV have been considered. As expected, according to the simple model given by Eq. 15, $C_{dp}$ is almost independent of both $E_g$ and $V$ (at both reverse and forward bias). As for $C_{df}$, there is a large difference between the two cases. In addition, the numerical results fit very well to the theoretical model given by Eq. 17. The large dependence of $C_{df}$ on $E_g$ comes from $n_i$ in Eq. 17.

These results show a larger depletion capacitance $C_{dp}$ in comparison to the diffusion capacitance $C_{df}$ even at forward bias, in remarkable difference with conventional (3D) p-n junctions.

The modeled capacitances in this Section allow device simulations (restricted to the intrinsic behavior) in the small-signal AC regime. Notice that no coupling effects between gates have been considered, so the model developed cannot deal with generalized AC simulations where AC voltages are applied to the gates. On the other hand, modeling of the extrinsic (fringing) capacitances is beyond the main purpose of this work.

## V. MODEL BENCHMARKING

Given the flexibility of the here studied device and the proposed numerical simulator, we simulate a 2D FETs by setting $V_{g1} = V_{g2}$, and $L_{gap} \to 0$. This provides an additional way to assess the simulator via comparison with reported experiments. Fig. 8a shows the transfer characteristics of a MoTe$_2$ FET, with SiO$_2$ as gate dielectric, for different applied bias according to the experimental data reported in [37] together with numerical simulations, with satisfactory agreement. In our simulations, we have assumed a contact width (Au/MoTe$_2$) of 3 µm and a gate-voltage independent contact resistance of 5×10$^4$ Ω-µm as an extrinsic series resistance, which is acceptable for ohmic contacts Au/MoTe$_2$ at high carrier densities as reported in [37,51–53]. On the other hand, Fig. 8b shows the output characteristics of a MoTe$_2$ based GT 2D LPNJ, with HfO$_2$ as gate dielectric, after applying opposite voltages $-V_{g1} = V_{g2} = 5$ V, with acceptable matching between the experimental data [37] and numerical results when a total contact resistance equal to 1.5×10$^5$ Ω-µm is assumed in the simulations. There is a strong rectification behavior or diode-like response with an RF about 5.2×10$^4$, where $RF = I(V = 1V)/I(V = -1V)$, which could be strongly affected by increasing the contact resistance. Ideality factor from numerical simulation is 2 whereas the one extracted from the experimental curve is 2.1. The small difference can be attributed to the use of a simple SRH model for the net recombination rate and to the parameters here considered constant as the carrier mobility.

Although some experimental research have reported that VRH mechanism dominates the carrier transport in disordered inorganic semiconductor based transistors at low density of carriers [43–45], the benchmarking presented in Fig. 8a-b and other works like [9,11,24,48,54] suggest that the DD model is consistent with experimental data. Up to now there is no report known by the authors about the influence of hopping mediated by interface trap states in 2D vertical or lateral pn junctions. Some evidence of the effect of VRH on the gate-tunable electrical characteristics of the pentacene/MoS2 p-n diodes at low temperatures has been reported [55]. On the other hand, a large ideality factor value (>>2) seems to be a common feature of vdW heterostructure devices where VRH transport mediated by interface trap states is significant [55–59].

## VI. CONCLUSIONS

We have developed a physics-based model to explore the out-of- equilibrium properties of the gate tunable 2D lateral PN junction. The proposed model, based on the numerical solution of the 2D Poisson's equation coupled with 1D drift-diffusion current and continuity equations, allows us to consider non-idealities such as the effect of the interface states. Our results reveal a simple linear dependence of the depletion layer width on both the insulator thickness and the built-in potential/applied gate voltage ratio and insensitivity to the dielectric constant of the gate insulator. On the other hand, the J-V characteristics show a rectifying behavior with an ideality factor close to two, which is indicative of the recombination current dominance within the depletion layer over the diffusion current produced in the charge induced region, consistently with experimental observations reported elsewhere. Notably, the model reveals that the presence of interface states produces a peak in the rectification factor versus the applied gate voltage. To assess the model, we have benchmarked it against experimental data corresponding to a gate tunable MoTe$_2$ based *pn* junction. The present work might be useful to assess compact models and opens the door to a wider exploration of potential advantages offered by gate tunable doped 2DPNJs in terms of figures of merits.

## ACKNOWLEDGMENTS

This work was funded in part by the EU's Horizon 2020 R&D Program (GrapheneCore2 785219 and GrapheneCore3 881603), in part by the MINECO (TEC2015-67462-C2-1-R), and by the MCIU (RTI2018-097876-B-C21). This article has been partially funded by the European Regional Development Funds (ERDF) allocated to the Programa Operatiu FEDER de Catalunya 2014–2020, with the support of the Secretaria d'Universitats i Recerca of the Departament d'Empresa i Coneixement of the Generalitat de Catalunya for emerging technology clusters to carry out valorization and transfer of research results. Reference of the GraphCAT project: 001-P-001702.

## DATA AVAILAVILITY

The data that support the findings of this study are available from the corresponding author upon reasonable request.

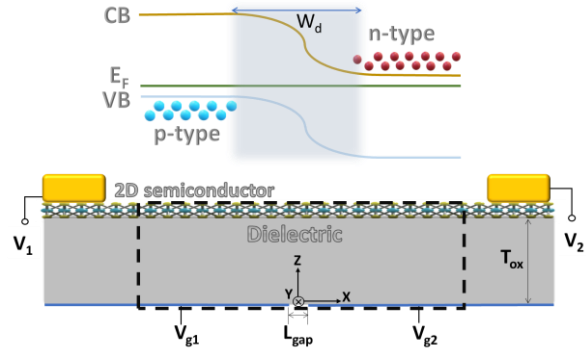

Fig. 1 Scheme of the physical structure of a general GT 2D LPNJ showing the applied voltages to each terminal. The region in grey corresponds to the depletion region and the dashed line represents the simulations window.

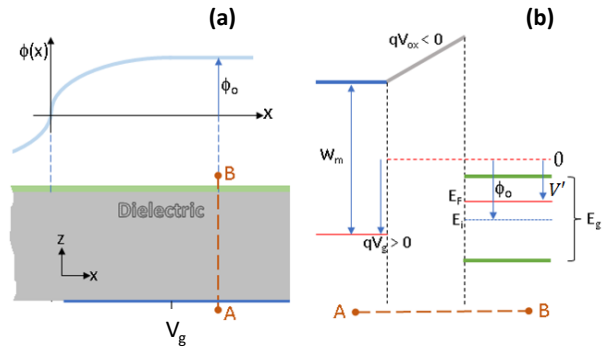

Fig. 2. (a) Electrostatic potential profile of the symmetric GT 2D LPNJ. (b) Band profile of the MOS structure along the AB line in (a).

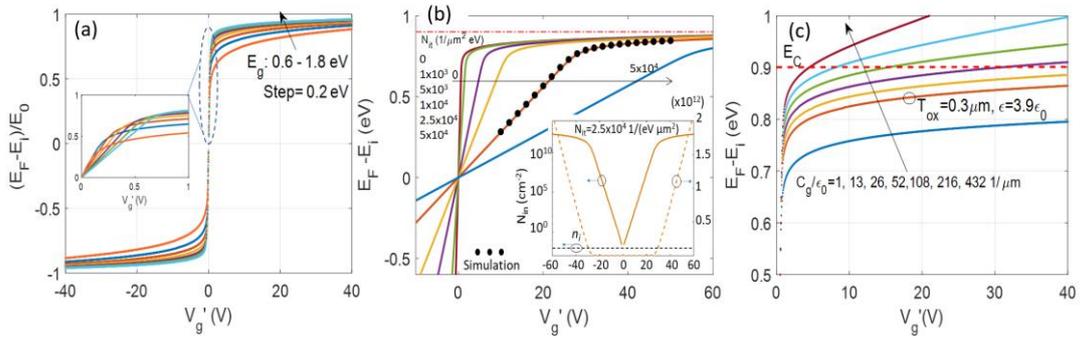

Fig. 3. (a) Normalized Fermi energy level as a function of $V_g'$ for several values of $E_g$ with $\epsilon = 3.9\epsilon_0$, $T_{ox} = 300$ nm and $N_{it} = 0$. (b) Fermi energy level as a function of $V_g'$ for several $N_{it}$ with $E_g = 1.8$ eV (MoS2), $\epsilon = 3.9\epsilon_0$ and $T_{ox} = 300$ nm. Inset: $N_{in}$ vs $V_g'$ for $N_{it} = 2.5E4 \ 1/(eV \ \mu m^2)$ in both linear and logarithmic scales. Solid black dots represent numerical solution of the 2D Poisson's equation (c) Fermi energy level as a function of $V_g'$ for several $C_g$ with $E_g = 1.8$ eV and $N_{it} = 0$. Red dotted lines in (b) and (c) represent the bottom edge of the conduction band.

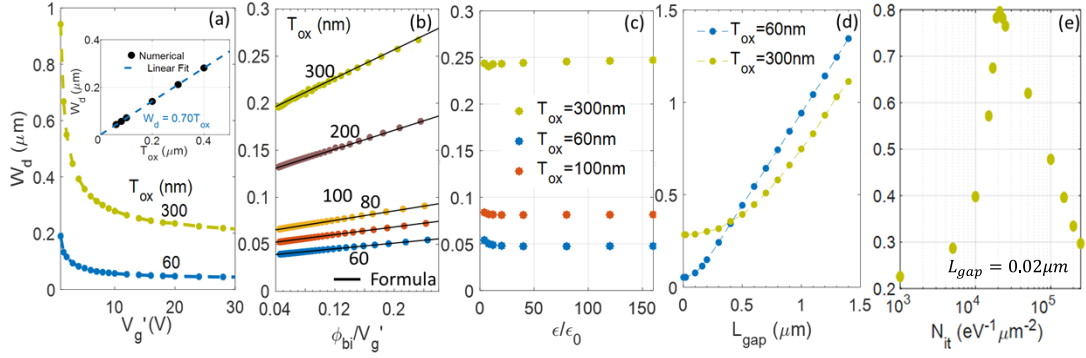

Fig. 4. Behavior of $W_d$ for symmetric GT 2D LPNJs ($V'_g = -V_{g1} = V_{g2}$) at thermal equilibrium with $E_g = 1.8$ eV (MoS$_2$), $\epsilon = 3.9\epsilon_0$ and $L_{gap} = 0.02$ µm. (a) $W_d$ vs $V'_g$ for two values of $T_{ox}$. Inset: $W_d$ vs $T_{ox}$ for a fixed value of $\phi_{bi}/V'_g = 0.08$. (b) $W_d$ vs $\phi_{bi}/V'_g$ for several values of $T_{ox}$. Black lines correspond to Eq. 10 and symbols to the simulations. (c) $W_d$ vs $\epsilon$ for several values of $T_{ox}$ at $V'_g = 10$ V. (d) $W_d$ vs $L_{gap}$ for two values of $T_{ox}$ at $V'_g = 10$ V. (e) $W_d$ vs $N_{it}$ for $T_{ox} = 300$ nm at $V'_g = 20$ V.

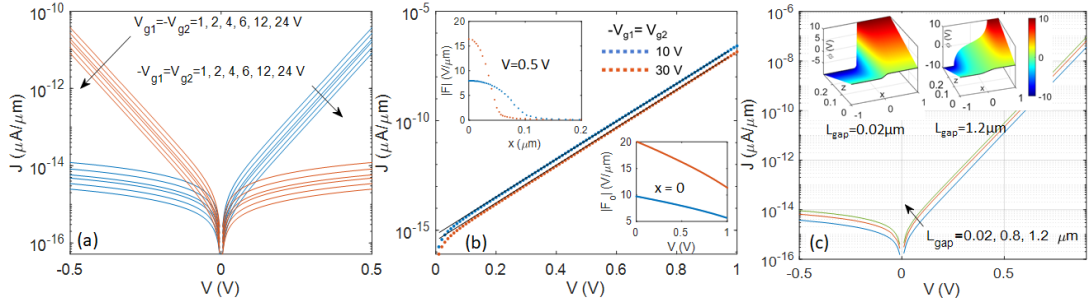

Fig. 5. J-V characteristics for symmetric GT 2D LPNJs with $E_g = 1.8$ eV (MoS$_2$), $\epsilon = 3.9\epsilon_0$, $T_{ox} = 300$ nm and $L_{gap} = 0.02$ µm. (a) J-V curves for several values $V'_g$. Blue and red lines correspond to *pn* and *np* junctions, respectively. (b) J-V curves at forward bias with $V'_g = 10$ V and 30 V. Symbols correspond to simulations and black solid lines to Eq. 11. Top inset: Electric field (F) profile at $V = 0.5$ V. Bottom inset: $F_o$ vs $V$. (c) J-V curves for several values of $L_{gap}$ at $V'_g = 10$ V. Inset: 2D electrostatic potential in equilibrium for two different values of $L_{gap}$.

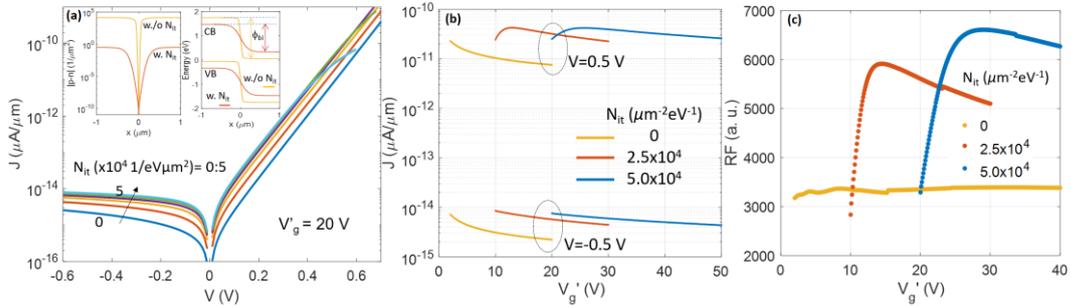

Fig. 6. (a) J-V curves for a GT 2D LPNJ with several values of $N_{it}$ polarized at $V'_g = -V_{g1} = V_{g2} = 20$ V. Inset: induced carrier density (left) and bands diagram (right) with and without presence of interface trapped charge (b) J-$V'_g$ curves for three different values of $N_{it}$ at reverse and forward biases. (c) Rectifying factor (RF) vs $V'_g$. Device parameters assumed are: $E_g = 1.8$ eV (MoS$_2$), $\epsilon = 3.9\epsilon_0$, $T_{ox} = 300$ nm and $L_{gap} = 0.02$ µm.

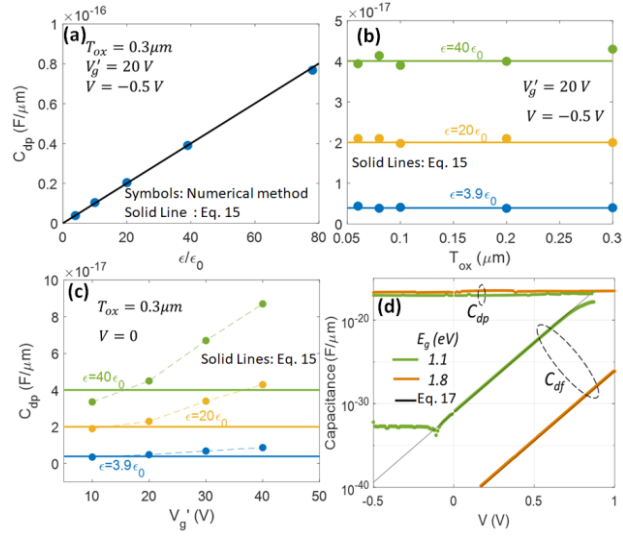

Fig. 7. Depletion and diffusion capacitances for the GT 2D LPNJ. Symbols correspond to simulations (a) $C_{dp}$ vs $\epsilon$. (b) $C_{dp}$ vs $T_{ox}$ for different $\epsilon$. (c) $C_{dp}$ vs $V_g'$ for different $\epsilon$. (d) $C_{dp}$ vs $V$ and $C_{df}$ vs $V$ for different band gap $E_g$.

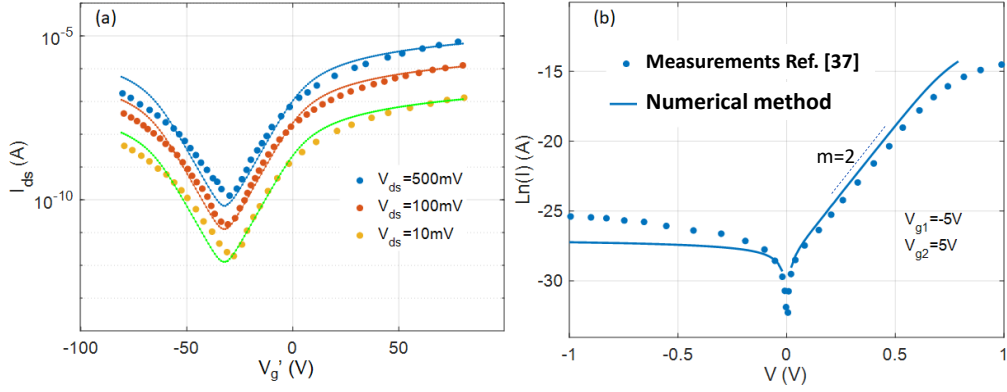

Fig. 8. (a) Transfer characteristics of the MoTe$_2$ FET. Symbols are experimental measurements and solid lines are simulations. Device parameters used are: $E_g = 1.0$ eV (MoTe$_2$), $T_{ox} = 300$ nm, $\epsilon = 3.9\epsilon_0$ (SiO$_2$), $L_{gap} = 0.02$ μm, $N_{it,A} = 2.8 \times 10^4$ eV$^{-1}$ μm$^{-2}$, $N_{it,D} = 7.86 \times 10^4$ eV$^{-1}$ μm$^{-2}$, $n_s = 1.23 \times 10^7$ μm$^{-2}$ (b) I-V curve for a MoTe$_2$ based GT 2D LPNJ. Device parameters used are: $E_g = 1.0$ eV (MoTe$_2$), $T_{ox} = 20$ nm, $\epsilon = 25\epsilon_0$ (HfO$_2$), $L_{gap} = 0.213$ μm, $N_{it,A} = N_{it,D} = 1.83 \times 10^5$ eV$^{-1}$ μm$^{-2}$.